\documentclass[final,5p,twocolumn]{elsarticle}
                                    
\usepackage{amsmath}             
\usepackage{amssymb}
\usepackage{pdfpages}
\usepackage{comment}
\usepackage{graphicx}          
\usepackage{gensymb}
\usepackage{float}

\journal{Photoacoustics}

\begin{document}

\begin{frontmatter}

\title{Compensating for visibility artefacts in photoacoustic imaging with a deep learning approach providing prediction uncertainties}

\author[lip]{Guillaume Godefroy\corref{email1}}
\author[lip]{Bastien Arnal}
\author[lip]{Emmanuel Bossy}
\address[lip]{Univ. Grenoble Alpes, CNRS, LIPhy, 38000 Grenoble, France}                                             
\cortext[email1]{Corresponding author: guillaume.godefroy@univ-grenoble-alpes.fr}
                                      
\begin{abstract}                         
Conventional photoacoustic imaging may suffer from the limited view and bandwidth of ultrasound transducers. A deep learning approach is proposed to handle these problems and is demonstrated both in simulations and in experiments on a multi-scale model of leaf skeleton. We employed an experimental approach to build the training and the test sets using photographs of the samples as ground truth images. Reconstructions produced by the neural network show a greatly improved image quality as compared to conventional approaches. In addition, this work aimed at quantifying the reliability of the neural network predictions. To achieve this, the dropout Monte-Carlo procedure is applied to estimate a pixel-wise degree of confidence on each predicted picture. Last, we address the possibility to use transfer learning with simulated data in order to drastically limit the size of the experimental dataset.
\end{abstract}


\end{frontmatter}

\section{Introduction}
Photoacoustic (PA) imaging is an emerging biomedical modality based on the generation of acoustic waves by light absorption. This modality is promising, as it enables imaging at large depths with high spatial and temporal resolution, and can provide images of the optical absorption \cite{beardPA} with specific molecular contrast which can be enhanced by spectroscopy.\\

In conventional PA imaging, a short nanosecond laser pulse is sent into the medium and the emitted ultrasonic waves are collected by a conventional ultrasound (US) probe. At the US propagation time scale, the object illumination is quasi instantaneous as the speed of light is several orders of magnitude higher than the speed of sound, resulting in the emission of strongly coherent acoustics waves \cite{guospec}. These waves interfere constructively or destructively depending on the structure of the object, often leading to two well-known artefacts on the reconstructed image: the limited bandwidth and the limited view artefacts \cite{deanspeck}.
With a resonant detection bandwidth, when an object larger than the acoustic central wavelength of the transducer is illuminated, the strong low frequency component of the PA signals is filtered out. With a limited view (limited detection aperture) , for a structure elongated along
the axis of the probe, the PA waves interfere constructively perpendicularly to the probe but mostly destructively throughout the elongation. As a result, very few signals are collected in this case by linear or matrix array probes due to their limited angular view. Both type of artefacts will further be referred to as the visibility problem in this paper.\\

The limited view problem has been addressed in several studies. The most intuitive approach is to either rotate the object relatively to the probe \cite{krugerroto} (or vice versa \cite{yangrotp}) or use ring shaped transducer arrays \cite{xiacyl}\cite{deanspeck} in order to cover all angles. However, a clinical implementation would benefit from a handheld real-time system as currently used in ultrasound imaging. Other approaches rely on the introduction of a spatial modulation of optical absorption of the sample, either using injection of sparse absorbing particles \cite{deanloc}, by a modulation of physical properties \cite{wanghea}, or by computing statistical properties of the PA signal generated by fluctuating sources in the medium \cite{chaignespec}\cite{vilov2020unified}. Fluctuations of the PA signals can be produced by random optical speckle pattern illuminations \cite{chaignespec} or by flowing red blood cells \cite{vilov2020unified}, naturally present in the blood vessels. Nevertheless, these methods require long acquisition times in order to get significant statistical properties, and therefore have a poor temporal resolution.\\

In this work, a deep learning approach is proposed to overcome the visibility problem and improve the image quality in a real-time single shot configuration. A neural network can be viewed as an algorithm composed of many parameters, called weights, designed to compute  input data into a desired form of output \cite{bengiodl}. This algorithm is trained over multiple examples to obtain the best representation of the studied phenomenon. After training, the network transforms an input raw image into an output image that is expected to resemble the (unknown) ground truth. The training set consists of multiple raw data/ground truth pairs that will be used to optimize the weights of the network.\\

Convolutional neural networks (CNN)are amongst the most popular category of deep learning algorithms (DLA)~\cite{lecunconv} , and have reached state of the art performances in several imaging problems including segmentation \cite{longseg}, classification \cite{russclass}, artefacts removal \cite{dongart} or denoising \cite{xieden}. CNN have been introduced recently in biomedical imaging, showing impressive results in various tasks \cite{akkusmri}\cite{jinct}\cite{liuus}. Over the past two years, a few groups started investigating deep learning applied to PA imaging for several purposes including direct reconstruction of the initial pressure \cite{waibelinitpre}, handling artefacts coming from sparse data \cite{hauptmannspars}\cite{antholzerspars}\cite{davoudispars}, reflection artefacts removal \cite{allman2018deep}\cite{shan2019accelerated}, point source localization \cite{reitersource}\cite{allmansource} and quantitative measurements \cite{caiquan}\cite{kirchnerquant}. The correction of the limited bandwidth problem was also investigated on very simple objects \cite{guttaband}. Some of these studies \cite{guandens}\cite{antholzerspars}\cite{davoudispars} showed that deep learning  can also reduce the limited view artefacts although results were either numerical or obtained with non-conventional imaging devices.
A linear array was used in experiments
\cite{kim2020deep} but a ground truth was missing to assess the success of the approach. Finally, in most of the cited studies, experimental results were predicted from models trained only on simulation data.\\

In this work, we focus on the correction of the whole visibility problem, induced both by the limited view and limited bandwidth of a conventional linear US probe. The originality of our approach resides in the design of a dedicated model object and a method to create an experimental training dataset. The method is used to  assess the capacity of a neural network to remove these artefacts on experimental images that were not used during the training. In this study, a ground truth is known also for the test set, which consisted of some of those unseen images. Thus, evaluation of the quality of the reconstruction can be performed. We point out that this study is not designed to produce quantitative PA images, as our ground truth does not directly represent the optical absorption, and focuses on providing morphological images. As a consequence, it can currently not be applied for quantitative imaging including multispectral investigations. Moreover, our study is limited to a given class of objects, and an investigation of the ability of the network to generalize to other classes of objects was out of the scope of this work. A preliminary discussion on generalization is provided in the supplementary materials. \\

Despite the impressive performances of DLA to reconstruct PA images, errors can be made by the algorithm which may misinterpret the data. This is one of the main limitations of neural network approaches in the medical field: the lack of confidence in the results. In this work, we estimate the uncertainty in our prediction through a Bayesian machine learning framework. We followed the approach proposed by Ghahramani and Gal \cite{gadropout}, referred to as Monte Carlo dropout (MC dropout), which has been recently applied for phase imaging \cite{xuphase}. Uncertainty estimation using a Bayesian framework has already been studied in PA imaging to reduce artefacts induced by approximation of reconstruction parameters \cite{sahlstrom2020modeling}\cite{tick2019modelling}. A specific deep learning approach has also been developed to estimate uncertainty of the optical parameter estimation in quantitative PA imaging \cite{grohl2018confidence}. Our case is different, as we do not want to take into account approximation of some parameters, but estimate the uncertainty linked to the deep learning process. To do so, our CNN is converted into a Bayesian neural network to introduce randomness in the prediction process, which makes the prediction no longer deterministic: the model will predict different outputs for the same input. Then, for a given input, several outputs are generated and are interpreted as samples of a probabilistic distribution, from which parameters can be estimated, such as the mean value and the confidence measure. The uncertainty estimation provides positions of invented and poorly reconstructed structures. This estimation is very useful for real-time navigation as a feedback for the user, who may eventually choose to display only the reliable parts of the images. However, we point out that this estimation does not provide information for structures missing in the reconstruction, and is therefore not a measurement of the fidelity. 
\\

\begin{figure*}[h!]
\centering
\includegraphics[scale=0.68]{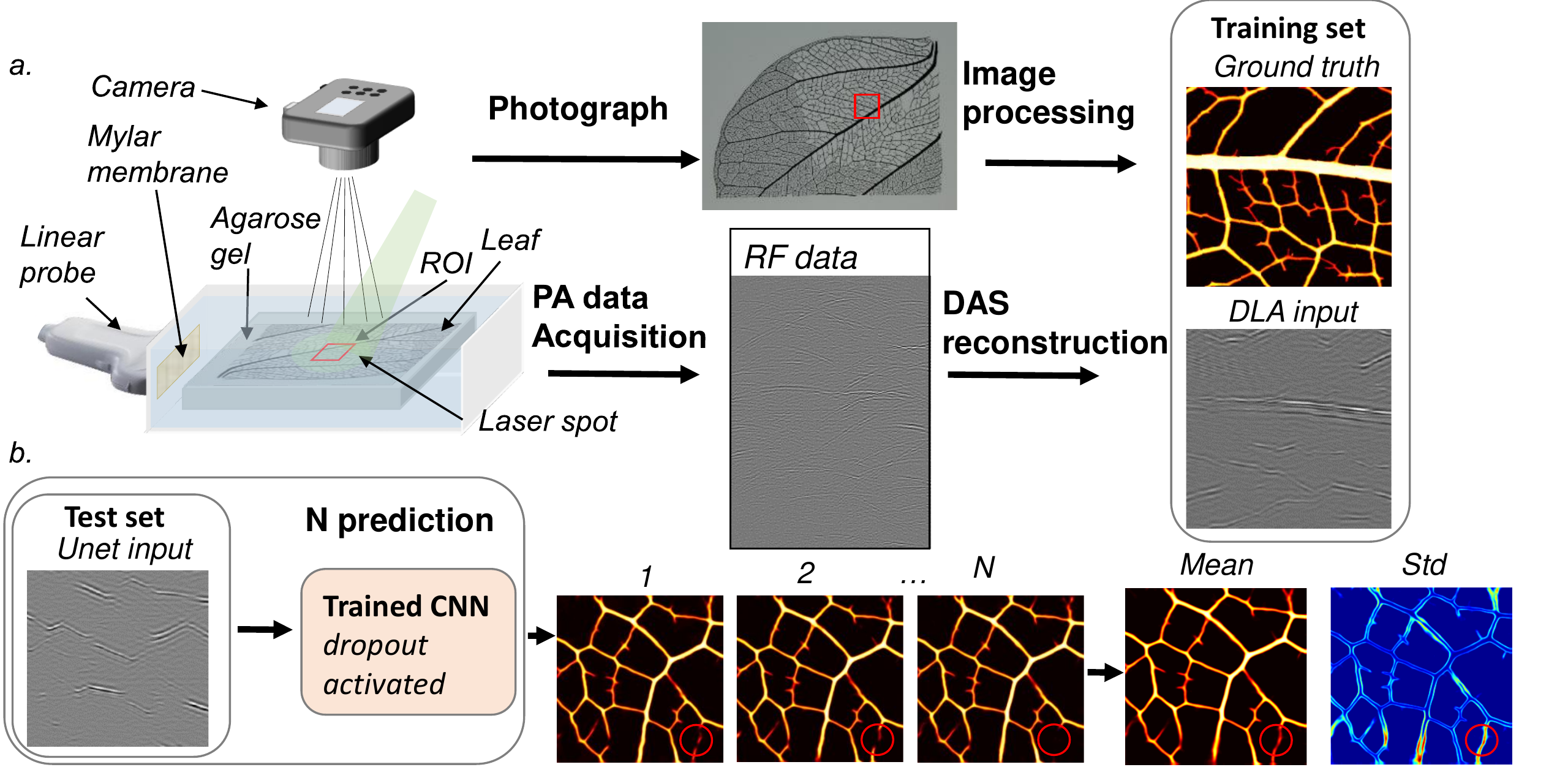}
\caption{
\textbf{a, }Creation of the experimental training set. A linear probe is coupled to a water tank containing the leaf, through a window composed of a tight Mylar membrane. The leaf is in the imaging plane of the probe. The laser beam is shined from the top and the RF signals are acquired. A photograph of the leaf was previously taken. The mBF PA image of the ROI is reconstructed and the photograph is processed to extract the same area.\textbf{ b, }Uncertainty prediction: Several images are generated using the same input. The mean and the standard deviation (std) of these samples are estimated pixel by pixel. The prediction is unstable in the marked area, resulting in a high std.}
\label{fig1}
\end{figure*}

We also study the DLA performance over different input data types. Usually, a conventionally reconstructed image is used. This prior reconstruction is obtained by applying delays and summation (DAS) on the Hilbert transform of the radiofrequency (RF) signals. This operation produces a complex image whose modulus is computed to form the demodulated beamformed (dmBF) image, which is the one displayed for the end user. 
While the input of the DLA for PA image reconstruction is usually the dmBF image, we choose to train our network with the modulated beamformed (mBF) image. The mBF image (Fig. \ref{fig1}.a) is obtained by applying DAS directly on the real-valued RF time signals. Consequently, the mBF image is modulated by the impulse response of the transducer, resulting in axial oscillations. While the mBF image represents the object less faithfully than the dmBF image (because of the probe oscillations), we show that it carries more information that the DLA can exploit. The two approaches are compared in Supplementary Materials. \\

Finally, we investigate the design of the training sets. Indeed, processing experimental data with CNN trained solely on simulated data seems to produce poor reconstruction \cite{davoudispars} which we confirm here (Supplementary Materials), while constructing a large experimental dataset is complex and time consuming. We varied the relative sizes of the combined experimental and simulated datasets and observed its impact on the reconstruction performances.

\section{Material and methods}

\subsection{Conventional reconstruction methods}
For comparison purposes, conventional DAS images (dmBF) and L2 deconvolution images are provided. DAS is fast and robust whereas deconvolution methods are more computational and more complex to implement since the knowledge of the point spread function of the system and regularization are necessary.
Here, image deconvolution is achieved using a least-square minimization approach with a L2-regularization penalty term. It was performed by a fast iterative shrinkage thresholding algorithm (FISTA) \cite{beckfista}. The inversion is defined as:
\begin{equation}
 \hat{X}=\underset{X}{argmin}     \frac{1}{2}\|Y-AX\|^2_2+\alpha^2  \|X\|_2       
\end{equation}
$\hat{X}$ is the expected reconstructed object, $X$ the object at each iteration, $Y$ represents the RF signals and $A$ is the propagation matrix, containing the imaging system response at each point of the reconstruction grid.
$\alpha$ is the regularization parameter, tuned heuristically by visually comparing  the reconstructed image with the ground truth.

\subsection{Creation of the experimental dataset}

A model of leaf skeleton was chosen as imaging sample (see Fig. \ref{fig1}.a). This model has been used in previous studies as it provides a branching structure qualitatively similar to that of a vascular network, and produces conventional images with similar visibility artefacts \cite{hojman2017photoacoustic}.
To obtain a sufficient photoacoustic signal, the leaves veins were tainted with black ink and the limbs were dissolved by chemical treatment. The smallest veins of the leaves are finally manually cut out to remove unresolvable details. Each pair of the dataset consisted of a mBF image (input of the network) and the corresponding photograph (ground truth) of a 5.12$\times$5.12 mm$^2$ patch of the considered leaf.\\

As shown in Fig. \ref{fig1}.a, the leaf is maintained inside a horizontal plane of an agarose gel, which stands within a tank filled with degassed and deionized water. Through a side window composed of a frame tightening a Mylar membrane, an ultrasonic transducer array (15.6 MHz central frequency, L22-8v, Verasonics, USA), connected to a multi-channel acquisition system (Vantage 256 High Frequency, Verasonics, USA) is coupled to the water tank with echographic gel. Thus, the leaf is in the imaging plane of the probe. It is illuminated from the top with 5 ns laser pulses at 10 Hz repetition rate \((\lambda = 532\) nm), produced by a frequency-doubled Nd:YAG laser (Surrelite, Continuum, USA). For each laser shot, PA signals are acquired and mBF images are reconstructed using DAS assuming a homogeneous medium with a speed of sound of 1500 m.s\(^{-1}\), neglecting the presence of the agarose gel. To obtain several independent samples from a same leaf, the leaf is  mechanically translated respectively to the probe and the light source. \\

We define our "ground truth" images as photographs of the leaf taken with a CMOS camera (X-E2, Fujifilm, Japan). These photographs are converted to grey scale (8 bits) and pixels below a threshold are set to 0 to suppress background noise. A registration between the PA image and the corresponding photograph is needed. The magnitude of the transformation to apply to co-register the two images (decomposed as rotation, translation and scaling) were found automatically by maximizing the correlation coefficient between the PA image (the reference) and the transformed photograph. 593 pairs of images from two leaves are obtained, split between the experimental training set and the experimental validation set  with respectively 500 and 93 pairs. The validation set is used during the training to assess the optimization process is over. An experimental test set of 15 pairs is then constituted from two different leaves. It is used to evaluate the performance of our approach. 

\subsection{Creation of the simulation dataset}

The two photographs used as ground truth for the experimental training set were also used to create the simulation dataset. Data augmentation is applied on those photographs to increase the dataset size by applying rotations, mirror transformations, horizontal or vertical shears, and center expansions or compressions. Then, 1$\times$1 cm images are extracted to compute  their PA signals. \\
The method used to simulate PA imaging is described in our previous work \cite{vilov2020super}. In brief, the imaging system response is experimentally measured for a single source at one spatial location and the synthesis of the RF signals coming from a whole object is obtained by summing the contributions of each pixel of the object. The object is then reconstructed with DAS as for the experimental data. The medium is assumed to be homogeneous with a constant speed of sound of 1500 m.s\(^{-1}\).
DAS is then applied to reconstruct an mBF image of 5.12$\times$5.12 $mm^2$ area and the photograph is cropped to match the dimensions.  Propagating PA waves from a larger area (1$\times$1 $cm^2$) than the one viewed by the network (5.12$\times$5.12 $mm^2$) enable to take into account the presence of surrounding structures which can affect reconstruction during experiments. A series of 1400 pairs of images are obtained. Around five days are needed to compute the dataset on Matlab with a desktop computer.

\subsection{Deep learning framework}
UNET \cite{ronnunet} is a widely used CNN in the medical field. A slightly modified architecture, presented in supplementary materials is implemented with the open source libraries Tensorflow and Keras. Dropout \cite{sridropout} and batch normalization \cite{ioffebn} layers are added to limit over-fitting of the model. The last layer contains only one filter instead of two in the original version, as the expected output is a single image. The last activation function is also suppressed as the prediction is no longer a binary image. It is worth mentioning that several modifications supposed to improve the result including skip connections between input and output \cite{akkusmri}, residual blocks \cite{diakogiannis2020resunet} and fully densely connected blocks \cite{guandens} have been tested without significant improvement of the prediction. The cost function is the classical mean squared error, and an Adam optimizer is used with a learning rate of $5.10^{-4}$ and momentum of 8 \cite{kingmaadam} with batch sizes of 8 images. An early stopping approach based on the validation loss was chosen to limit under- and over-fitting \cite{prechearly}. 
The prediction phase must be random to model uncertainty. In the MC dropout approach, noise is injected in the model by activation of the dropout layers (dropout rate of 50$\%$) both during training and prediction. In this study, 20 inferences are generated from forward passes through the model with a different dropout mask. The different resulting predictions allow to further estimate the distribution mean and its standard deviation which gives a map of uncertainty (see Fig. \ref{fig1}.b). The training and the evaluation of the network, composed by around 30,000,000  neurons, are performed on a NVIDIA Quadro P2000. Around 50 minutes are needed for the training on the simulation set and 30 minutes on the experimental set. 

\subsection{Quantitative assessment of the network performance}

As mentioned previously, the same leaves are used to create the simulation and the experimental dataset.
It means that from the same object (i.e. an area of the leaf), we will be able to obtain either the experimental RF signals or the simulated one. For comparison purpose, the reconstructed objects shown in the figures are the same for both simulations and experiments. A third example is used for the MC dropout results, the estimated uncertainties of the two previous examples being described in the supplementary materials.
All images are normalized by their maximum and represented with a colorbar from 0 to 1, since quantitative information is not expected.  \\

To evaluate the accuracy of a trained neural network, the normalized 2D cross-correlation \cite{yooncc} (NCC) and the scaled and shifted structured similarity index (sSSIM) are computed between each output and ground truth. The first one uses local sum to normalize the cross-correlation for feature matching. SSIM \cite{wangssim} is a widely-used metric to evaluate the perceived quality of an image. It is computed over several small windows of the image, quantifying the structure, contrast and intensity similarities. The sSSIM \cite{schwabssim} is used for obtaining a scaled and unbiased score which was not disadvantaging for the other reconstruction methods. For an overall performance estimation of the network, the mean and standard deviation values among the test set are presented. On each result, we computed an uncertainty map with the MC dropout method, which only require a single-shot acquisition. For comparison purposes, as an alternative way to assess the local variability in the reconstructions, we also computed an uncertainty map from the standard deviations of the reconstructions of 50 acquisitions of the same sample. In this case, the variability comes from the experimental noise while the CNN remain deterministic. We also computed the absolute truth error of the reconstruction defined as the difference between the ground truth and the predicted image.

\section{Results}
\subsection{Simulation results}
\begin{figure}[H]
\centering
\includegraphics[scale=0.5]{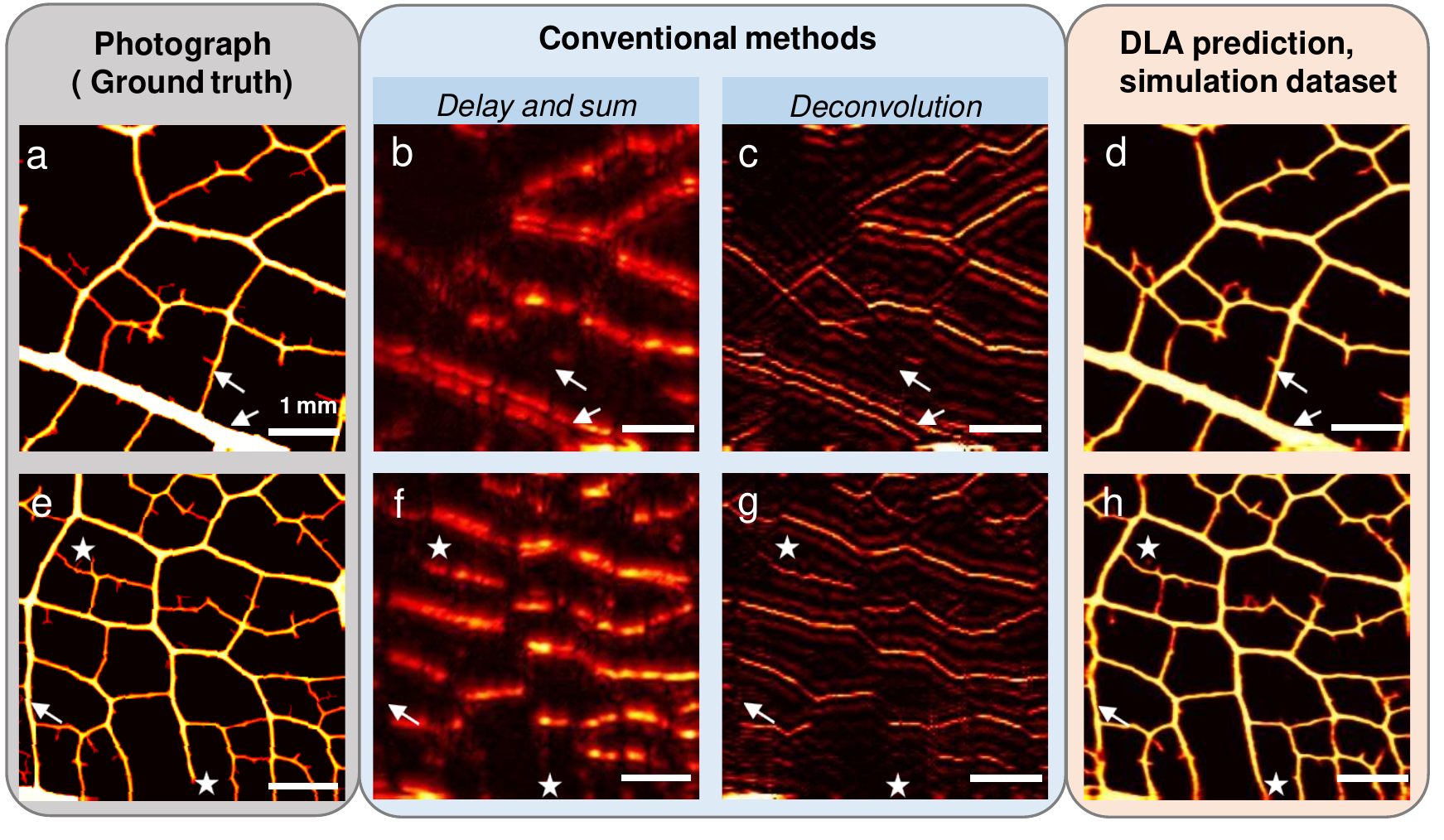}
\caption{Results on the simulated test set built from simulated RF signals, two examples. \textbf{a, e, }Ground truth: photograph of the object.\textbf{ b, f, }dmBF image, delay and sum.\textbf{ c, g, }L2-regularized deconvolution.\textbf{ d, h, }Prediction of the deep learning algorithm.
}
\label{fig2}
\end{figure}

Fig. \ref{fig2} shows a comparison between the reconstructed image from simulated data provided by the DLA (d,h) (trained with simulated data), the dmBF image (DAS, b,f), a L2-regularized deconvolution (c,g) and the photograph of the object (ground truth, a,e). L2 minimization and DAS clearly do not provide vertical structures, i.e. the structures elongated in the axis of the probe. Veins having inclinations beyond the detection aperture (typically beyond 45 degrees) are missing due to the limited view problem. The inside of the thicker vessel is also missing and the thickness of the thinner vessel is underestimated for DAS reconstruction and overestimated for the L2 minimization (arrows). In contrast, the deep learning reconstruction yields  an almost artefact-free reconstruction with errors located only on the smallest appendages resulting from the manual cutting, and on few vertical structures which are not completely recovered (stars on the images). 
The performances of the different reconstruction methods and their standard deviation, evaluated over the 15 pairs of the simulation test set with the metric described in 2.5 are shown in Tab.~\ref{tabl}. These numbers clearly confirm the qualitative visual impression:  when the DLA is used, the NCC and sSSIM are about three times higher compared to the simple DAS. Scores for the deconvolution method, not shown here, are on the same order than that of the DAS. 
\begin{table}[htbp]
  \caption{Quantitative measurement of reconstruction quality with the normalized  2D  cross-correlation and the scaled  and  shifted  structured  similarity  index}
   \label{tabl}
  \begin{tabular*}{\hsize}{lrrrrr}
\hline
&\multicolumn{2}{c}{Simulation} &\multicolumn{2}{c}{Experiment}\\
\hline
    & DAS & DLA &DAS &DLA \\
    \hline
        NCC & $0.31\pm.02$  & $0.89\pm.01$ & $0.44\pm.06$ & $0.80\pm.03$ \\
        sSSIM & $0.29\pm.02$ & $0.87\pm.01$ & $0.38\pm.05$ & $0.76\pm.03$  \\
        \hline
  \end{tabular*}
\end{table}
\subsection{Experimental results}
\begin{figure}[H]
\centering
\includegraphics[scale=0.5]{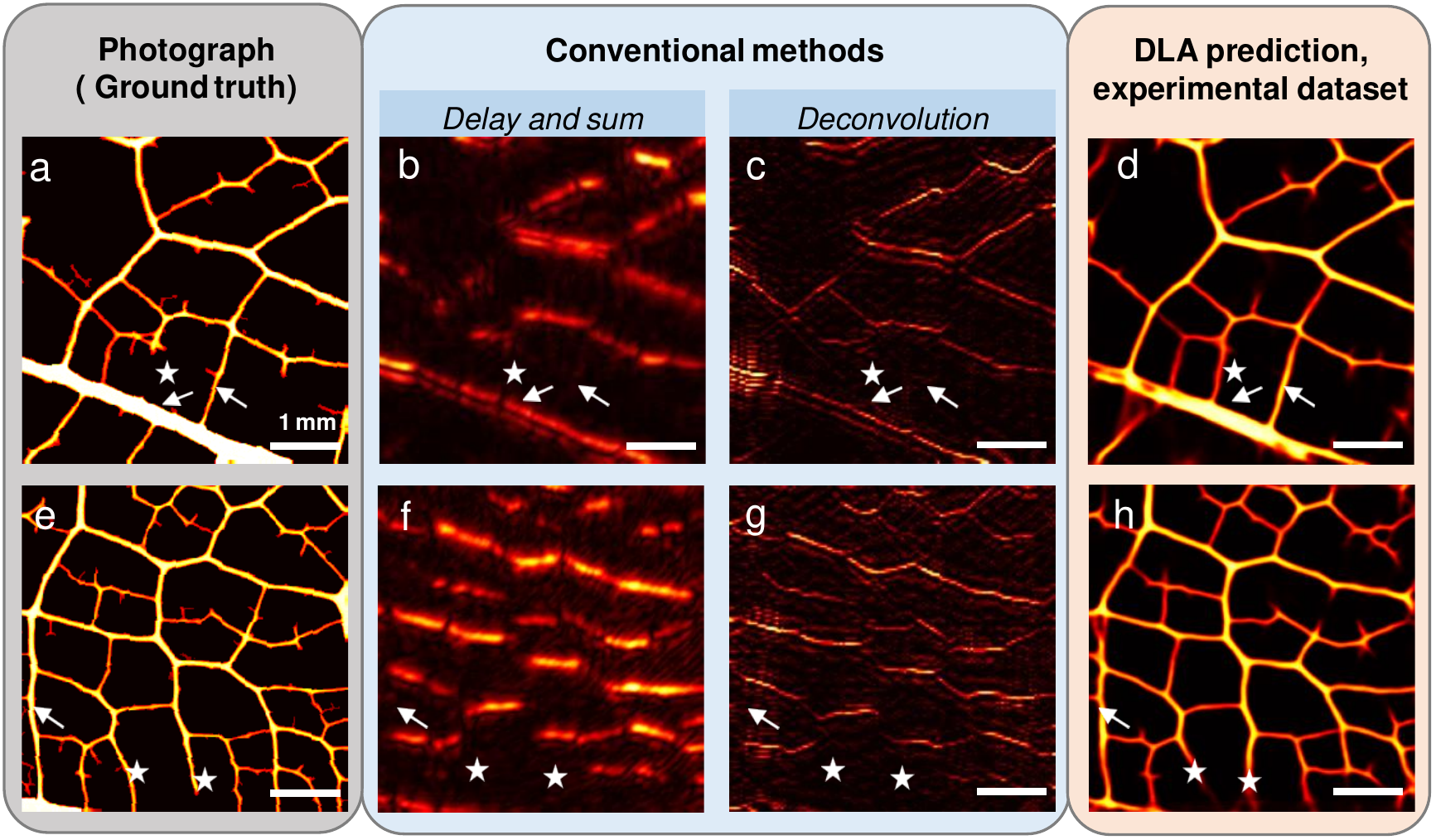}
\caption{Results on the experimental test set built from experimentally acquired RF signals, two examples. \textbf{a, e, }Ground truth: photograph of the object.\textbf{ b, f, }dmBF image, delay and sum.\textbf{ c, g, }L2-regularized deconvolution.\textbf{ d, h, }Prediction of the deep learning algorithm.}
\label{fig3}
\end{figure}

Fig. \ref{fig3} shows a comparison between the reconstructed image from the experimental data provided by the DLA (d,h) (trained with experimental data), the conventional DAS reconstruction (b,f), a L2-regularized deconvolution (c,g) and the photograph of the object (ground truth, a,e). Similarly, the DAS approach and the L2-minimization both fail to recover vertical structures as well as to provide a good rendering of the vein thicknesses by filling the inside of the thicker ones. In contrast, the DLA trained on the experimental data yields a reconstruction with most of the vertical structures recovered and a correct thickness of the veins (arrows). A few structures are again not recovered, and some mistakes occurred especially for the reconstruction of vertical veins (stars). The orientation is not always perfectly respected, and in some places, some veins appear when there are none in reality. 

Quantitative performances are shown in Tab.~\ref{tabl} where as for simulations, we observe a large improvement for the deep learning approach comparing to the DAS, although lower than simulation results. Both the sSSIM ($0.76$) and the NCC ($0.8$) are significantly enhanced. It may also be noted that the DAS performs better on experimental data than on simulation data.

\subsection{Uncertainty estimation}

\begin{figure}[H]
\centering
\includegraphics[scale=0.64]{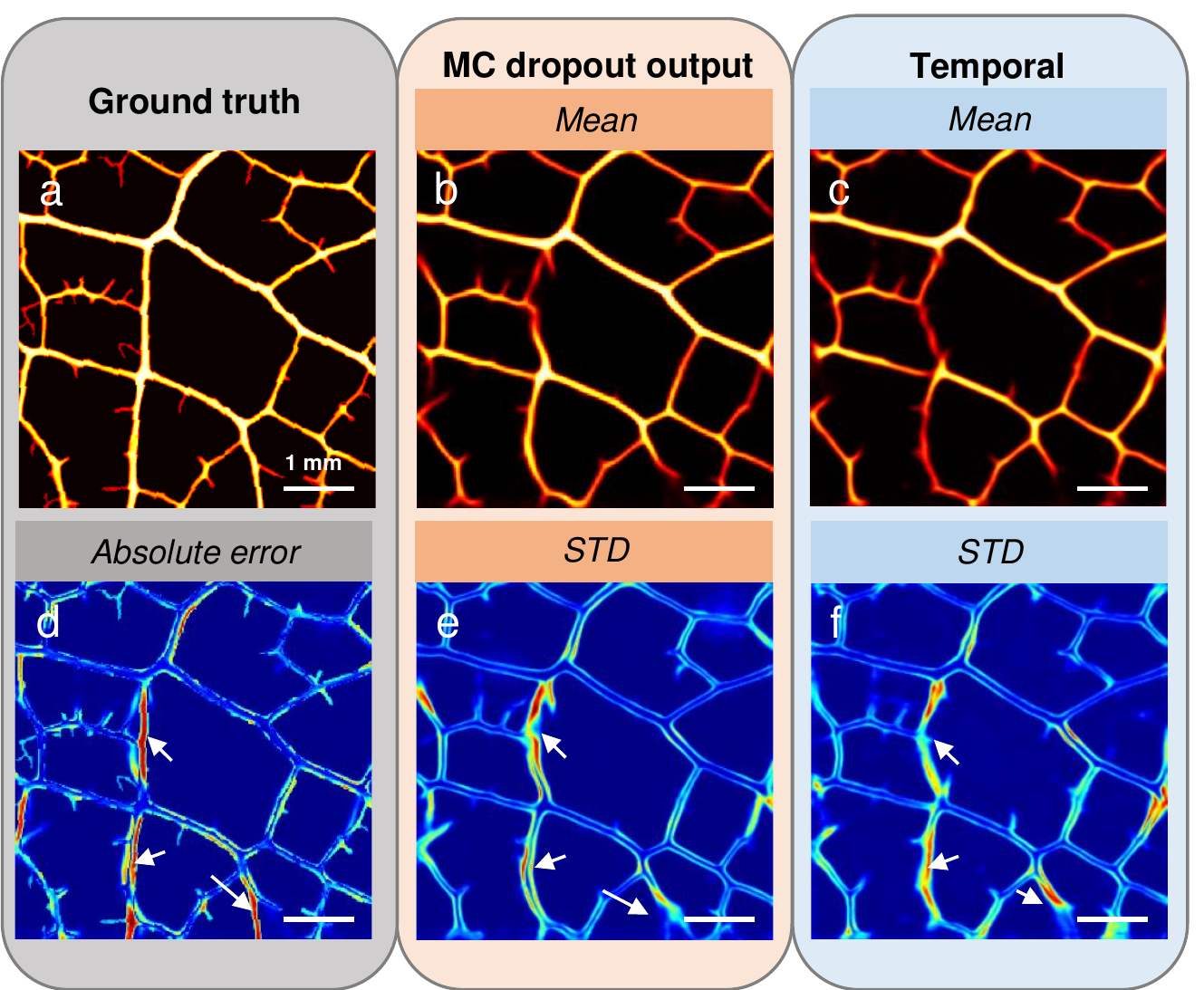}
\caption{Uncertainty estimation on predicted experimental image. 
\textbf{a, }Ground truth: photograph of the object.\textbf{ b, e, }Mean image and standard deviation of the object computed over 20 inferences generated from a unique acquisition with the deep learning algorithm, dropout activated.\textbf{ c, f, }Mean image and standard deviation of the object computed over prediction generated from several RF signals acquisition (experimental variability) with the deep learning algorithm, dropout disabled.\textbf{ d, }Absolute error between the ground truth (a) and the mean (b).}
\label{fig4}
\end{figure}

Results of the MC dropout procedure are presented in Fig. \ref{fig4}. A low standard deviation indicates a good robustness of the technique: the prediction remains stable over several realizations. Areas with high value are similar in the estimated uncertainty map (Fig. \ref{fig4}.e), the absolute error map (Fig. \ref{fig4}.d) and in the map of standard deviation from experimental noise (Fig. \ref{fig4}.f). Most of the errors in the prediction are captured like the small vessel at the bottom, that was not fully recovered, or the central one which was reconstructed but in a curved shape instead of a straight one (arrows).

\subsection{Impact of the pretraining and the size of the training set on the performance}
\begin{figure*}[h!]
\centering
\includegraphics[scale=0.9]{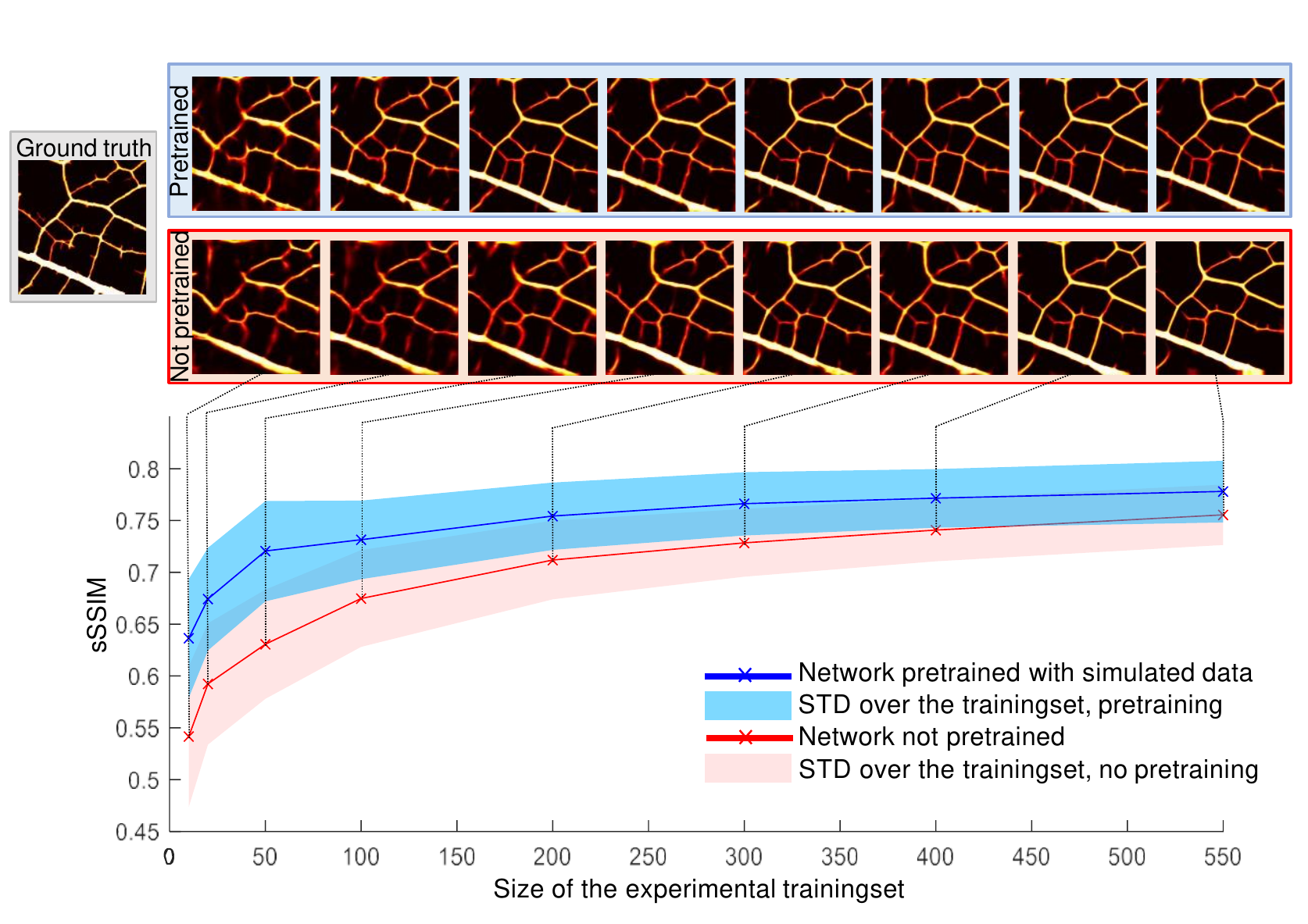}
\caption{Performance of a deep learning algorithm trained on an experimental dataset only (red) versus a deep learning algorithm pretrained on 1400 simulated data then trained on the same experimental dataset (blue), for different number of experimental data.}
\label{fig5}
\end{figure*}
In this part, the efficacy of a pretraining session with a simulation dataset is investigated as a means to improve the general performance and for reducing the size of the experimental training set. The uncertainty prediction was not studied in this configuration. We increase the training set size with unseen part of the two leaves used for the testset. The DLA was trained with experimental datasets of different sizes, from 10 to 550 pairs (the entire dataset). For each size, the training is repeated 30 times with, for each of them, a training and validation set composed of different pairs randomly chosen. This is needed to limit the influence of individual pairs on the training set size (for example, it is likely that 10 examples very close to the test set will provide a better prediction than 20 very distant ones), especially for small set sizes. The displayed sSSIM values are therefore an average over all the test sets from the 30 different realizations. To evaluate pretraining, we repeated this procedure with weights initialized by those obtained at the end of a training session on a simulation dataset composed of 1400 pairs. \\

The results are shown in Fig. \ref{fig5}. As expected, the performance increases with the experimental dataset size. Below 200 pairs, errors remain present and the veins thickness is not always faithfully represented. From 200 pairs, the image quality seems visually stable, although the sSSIM value still slowly increases. With pretraining (blue curve) convergence is faster. When the full dataset is used, pretraining only slightly increases the performance of the network (sSSIM improved from 0.76 to 0.78). For a smaller size such as 50 pairs, the score improvement is better (from 0.63 to 0.72). A reconstructed image comparable to the one obtained with the total experimental training set is almost reached from this experimental dataset constituted of 50 pairs. In this situation, a pretraining session therefore enables to decrease the size of the experimental training set by a factor 4.

\section{Discussion}

The algorithm trained with simulated data is able to produce images that are free from visibility artefacts, when applied to simulated test data. When trained with simulated data, the algorithm applied to experimental data however fails to provide images of good quality, as illustrated in the Supplementary Materials (see fig. S3).
When both training and prediction are made with experimental data, while a few errors may remain, most vessels are well recovered. A fundamental difference between simulation and experiment is the nature of the ground truth. In simulation, the ground truth exactly represent the distribution of absorbed light. In the experiments, the ground truth is a photographic 2D projection of a three dimensional absorption distribution. It then represents the integration of the optical absorption through the sample thickness (which may vary among leaf veins). Consequently, the photograph is not a quantitative representation of the sample absorption, and cannot therefore not be a quantitative ground truth. Thus, our method is not supposed to provide a quantitative reconstruction of the absorbed light, as the network is forced to learn from a 2D representation of a 3D object of finite and varying thickness. Given the nature of our ground truth, the purpose here was to demonstrate that the morphology of the sample could be retrieved free of visibility artefact. Quantitative ground truth would be required to provide quantitative information on the absorption coefficient, as needed for instance for spectroscopic applications. \\

The use of the mBF image as input of the network improves the performance both quantitatively and qualitatively compared to the dmBF image input (see supplementary materials, Fig. S2). In fact, the mBF image, 
although affected by the oscillatory impulse response to the ultrasound probe,
turns out to carry more information to be captured by the network.\\

The results show that errors are often located at the edges of the reconstructed image. Indeed, in these areas, less information about the surrounding structures is available. One way to limit these artefacts could be to reconstruct on a larger area and crop the edges. Most of the errors remaining on experimental images are located where the manual cutting was performed resulting in small appendices. These structures, which do not belong to the initial object, turned out to mislead the network which seeks to elongate them to join all the veins together. It is reasonable to think that the number of errors would have been lower on a more regular object. More broadly speaking, these results can be enhanced by improving the quality of the training set. \\

Nevertheless, potential errors in the reconstructed image, especially the invented structures are problematic for end users (clinicians, biologists...). The MC dropout approach proposed in this article helps locating most of them. Importantly, the estimated uncertainty remains low where reconstruction is correct, leading to a clear distinction of the suspicious areas: false alarms, which could mislead the clinician, are rare. It is worth mentioning that if the method helps locating the invented structures or incorrect reconstruction, (true positive), it is less efficient when capturing missed structure (false negative),  as illustrated in the supplementary materials, Fig. S4. This is understandable, as these errors are mostly related to a lack of information in the data. The map of the standard deviation over experimental realizations was computed to compare our result with the uncertainty map based on making the CNN stochastic. The areas showing high values in the two maps are co-located, showing that the MC dropout method with a single-shot acquisition provides an information similar to that resulting from noise induced variability. This feature is promising in the context of moving tissue imaging and real time navigation. We explored other methods that can provide uncertainty maps, such as Deep ensemble \cite{laksde} and Dropout ensemble \cite{xuphase}. In our study, the estimation of the mean image turned out to be superior for MC dropout according to the metrics presented previously. This difference could be explained by the required modification of the loss function for the other methods, involving a decrease of the overall performance. \\

To obtain the result of our study, our model was trained on an experimental dataset. However in clinical context, large experimental datasets may be complex to build. Using only simulated data to train a model and reconstruct experimental images would be ideal as simulation data can easily by produced. In our study, this approach turned out to  produce unsatisfactory results, as we illustrate in supplementary materials (Fig.S3): predictions on experimental data provided by a DLA trained on either simulated data or experimental one are shown, and show that many artefacts remain on the experimental images reconstructed from a DLA trained with simulation results. These results are also in agreement with observations made by Davoudi et al. \cite{davoudispars}. Although simulations are supposed to correctly model experiments, the nature of the ground truth for both training is very different for simulations and experiments in our case, which is likely the reason why training from experimental data provides much better results as compared to training with simulation data. Nonetheless, as shown in the previous section, the incorporation of simulated data through a transfer learning approach allows reducing significantly the size of the experimental dataset. The algorithm only needs to update its parameters with the difference between the simulations and the experiments, which is easier than learning the overall procedure. In the medical field, such a pretraining session could be useful for reducing the number of patients necessary to create a training set. \\ 

While our objective was limited to a proof-of-concept demonstration, several challenges must be taken into account in order to apply our method in a practical context.\\

The ground truth used for our proof of concept has to be replaced in practice by a ground truth that can be measured in a realistic environment (including deep inside tissue). Such a ground truth measurement could come from another imaging modality such as X-ray computer tomography (X-ray CT) or magnetic resonance imaging (MRI), which can accurately retrieve morphological information. They would however not provide a quantitative ground truth for the optical absorption, which would prevent the use of the method for  spectroscopic approaches.
A quantitative estimate of the ground truth can be obtained with a more sophisticated PA imaging device \cite{davoudispars}, or any method providing a visibility artefact free image proportional to the optical absorption. The training could be done with such a device free of visibility artefacts, to train a DLA to be applied on a simpler device. Quantitative reconstruction has been obtained with such an approach in the context of simulations~\cite{bench2020towards}\\

The influence of noise on RF signals should also be studied to assess the validity of our approach in a noisier environment. In our work, the signal to noise ratio (SNR) on RF signals is about 60. This value however represents the SNR of signals produced by horizontal structures, while this work mainly focuses on the reconstruction of vessels affected by the visibility problem, for which the signal is almost nonexistent. In addition, the background of our DAS images is polluted by clutter, an artefact located around the object originating from the lack of information for the reconstruction. The amplitude of the clutter is often higher than the one from vertical structures. The model situation considered in this work thus remains significantly challenging.  \\

Finally, the quality of the prediction is strongly influenced by the class of the object to reconstruct. The relative homogeneity of the studied dataset is one of the reason the DLA performs well: while very good predictions were made for leaves that were never seen by the network during the training, the unseen leaves were for the same species than the leaves used for the training. This is however a quite inherent limitation of deep learning approaches. In the supplementary information, we provide a preliminary investigation on the ability of our approach to generalize prediction to objects that do not belong to the class used for training: we tried to reconstruct the vessel structure available in the k-wave package (http://www.k-wave.org/) with the algorithm trained on the simulated dataset constructed with leaves. The predicted image, showed in supplementary materials (Fig. S5), is well reconstructed without artefact, suggesting there is no, or low overfitting in our approach, and that the network may generalize well. In a more general context, the capacity of a network to generalize is crucial and must be investigated for each particular situation.\\

Aside from increasing the quality of the reconstructed image, DLA offers other interesting features. Only 10 ms is needed to make a prediction using a regular graphic card which is much lower than the reconstruction time for the deconvolution method. Real-time reconstruction during user navigation could be achieved. Besides, once trained, the network does not need any parameters to be set by the user, unlike for deconvolution approach where the regularization parameter has to be chosen carefully and in a rather subjective way. 

\section{Conclusion}
The possibility of removing visibility artefacts with a neural network has been  demonstrated both in simulations and experiments on a model class of complex objects. Vertical parts of objects and the inside of large structures, missing on conventional reconstruction approaches, are recovered. These qualitative assessments are confirmed by quantitative metrics, which are far better for the DLA approach compared to 
conventional reconstruction methods. However, some errors may still be present in the reconstructed images, such as invented or poorly reconstructed structures as well as missing structures, although their number might be reducible by improving the experimental protocol. A MC dropout approach was proposed and successfully applied to identify invented and poorly reconstructed structures: high values on the generated uncertainty map are in agreement with high values on the true error map. Besides, it has been shown that pretraining the network with simulated data enables to reduce the size of the experimental training set by a factor of 4 while maintaining a similar quality of reconstruction.

\section{Funding Information}

This project has received funding from the European Research Council (ERC) under the European Union’s Horizon 2020 research and innovation program (grant agreement No 681514-COHERENCE).

\bibliographystyle{elsarticle-num}    
\bibliography{biblio}          

\begin{thebibliography}{10}
\expandafter\ifx\csname url\endcsname\relax
  \def\url#1{\texttt{#1}}\fi
\expandafter\ifx\csname urlprefix\endcsname\relax\def\urlprefix{URL }\fi
\expandafter\ifx\csname href\endcsname\relax
  \def\href#1#2{#2} \def\path#1{#1}\fi

\bibitem{beardPA}
P.~Beard, Biomedical photoacoustic imaging, Interface focus 1~(4) (2011)
  602--631.

\bibitem{guospec}
Z.~Guo, L.~Li, L.~V. Wang, On the speckle-free nature of photoacoustic
  tomography, Medical physics 36~(9Part1) (2009) 4084--4088.

\bibitem{deanspeck}
X.~L. De{\'a}n-Ben, D.~Razansky, On the link between the speckle free nature of
  optoacoustics and visibility of structures in limited-view tomography,
  Photoacoustics 4~(4) (2016) 133--140.

\bibitem{krugerroto}
R.~A. Kruger, W.~L. Kiser~Jr, D.~R. Reinecke, G.~A. Kruger, Thermoacoustic
  computed tomography using a conventional linear transducer array, Medical
  physics 30~(5) (2003) 856--860.

\bibitem{yangrotp}
D.~Yang, D.~Xing, S.~Yang, L.~Xiang, Fast full-view photoacoustic imaging by
  combined scanning with a linear transducer array, Optics express 15~(23)
  (2007) 15566--15575.

\bibitem{xiacyl}
J.~Xia, M.~R. Chatni, K.~I. Maslov, Z.~Guo, K.~Wang, M.~A. Anastasio, L.~V.
  Wang, Whole-body ring-shaped confocal photoacoustic computed tomography of
  small animals in vivo, Journal of biomedical optics 17~(5) (2012) 050506.

\bibitem{deanloc}
X.~L. Dean-Ben, D.~Razansky, Localization optoacoustic tomography, Light:
  Science \& Applications 7~(4) (2018) 18004--18004.

\bibitem{wanghea}
L.~Wang, G.~Li, J.~Xia, L.~V. Wang, Ultrasonic-heating-encoded photoacoustic
  tomography with virtually augmented detection view, Optica 2~(4) (2015)
  307--312.

\bibitem{chaignespec}
T.~Chaigne, B.~Arnal, S.~Vilov, E.~Bossy, O.~Katz, Super-resolution
  photoacoustic imaging via flow-induced absorption fluctuations, Optica 4~(11)
  (2017) 1397--1404.

\bibitem{vilov2020unified}
S.~Vilov, G.~Godefroy, B.~Arnal, E.~Bossy, Photoacoustic fluctuation imaging:
  theory and application to blood flow imaging, Optica 7~(11) (2020)
  1495--1505.

\bibitem{bengiodl}
Y.~Bengio, I.~Goodfellow, A.~Courville, Deep learning, Vol.~1, Citeseer, 2017.

\bibitem{lecunconv}
Y.~LeCun, L.~Bottou, Y.~Bengio, P.~Haffner, et~al., Gradient-based learning
  applied to document recognition, Proceedings of the IEEE 86~(11) (1998)
  2278--2324.

\bibitem{longseg}
J.~Long, E.~Shelhamer, T.~Darrell, Fully convolutional networks for semantic
  segmentation, in: Proceedings of the IEEE conference on computer vision and
  pattern recognition, 2015, pp. 3431--3440.

\bibitem{russclass}
O.~Russakovsky, J.~Deng, H.~Su, J.~Krause, S.~Satheesh, S.~Ma, Z.~Huang,
  A.~Karpathy, A.~Khosla, M.~Bernstein, et~al., Imagenet large scale visual
  recognition challenge, International journal of computer vision 115~(3)
  (2015) 211--252.

\bibitem{dongart}
C.~Dong, Y.~Deng, C.~Change~Loy, X.~Tang, Compression artifacts reduction by a
  deep convolutional network, in: Proceedings of the IEEE International
  Conference on Computer Vision, 2015, pp. 576--584.

\bibitem{xieden}
J.~Xie, L.~Xu, E.~Chen, Image denoising and inpainting with deep neural
  networks, in: Advances in neural information processing systems, 2012, pp.
  341--349.

\bibitem{akkusmri}
Z.~Akkus, A.~Galimzianova, A.~Hoogi, D.~L. Rubin, B.~J. Erickson, Deep learning
  for brain mri segmentation: state of the art and future directions, Journal
  of digital imaging 30~(4) (2017) 449--459.

\bibitem{jinct}
K.~H. Jin, M.~T. McCann, E.~Froustey, M.~Unser, Deep convolutional neural
  network for inverse problems in imaging, IEEE Transactions on Image
  Processing 26~(9) (2017) 4509--4522.

\bibitem{liuus}
S.~Liu, Y.~Wang, X.~Yang, B.~Lei, L.~Liu, S.~X. Li, D.~Ni, T.~Wang, Deep
  learning in medical ultrasound analysis: a review, Engineering (2019).

\bibitem{waibelinitpre}
D.~Waibel, J.~Gr{\"o}hl, F.~Isensee, T.~Kirchner, K.~Maier-Hein, L.~Maier-Hein,
  Reconstruction of initial pressure from limited view photoacoustic images
  using deep learning, in: Photons Plus Ultrasound: Imaging and Sensing 2018,
  Vol. 10494, International Society for Optics and Photonics, 2018, p. 104942S.

\bibitem{hauptmannspars}
A.~Hauptmann, F.~Lucka, M.~Betcke, N.~Huynh, J.~Adler, B.~Cox, P.~Beard,
  S.~Ourselin, S.~Arridge, Model-based learning for accelerated, limited-view
  3-d photoacoustic tomography, IEEE transactions on medical imaging 37~(6)
  (2018) 1382--1393.

\bibitem{antholzerspars}
S.~Antholzer, M.~Haltmeier, J.~Schwab, Deep learning for photoacoustic
  tomography from sparse data, Inverse problems in science and engineering
  27~(7) (2019) 987--1005.

\bibitem{davoudispars}
N.~Davoudi, X.~L. De{\'a}n-Ben, D.~Razansky, Deep learning optoacoustic
  tomography with sparse data, Nature Machine Intelligence 1~(10) (2019)
  453--460.

\bibitem{allman2018deep}
D.~Allman, F.~Assis, J.~Chrispin, M.~A.~L. Bell, Deep neural networks to remove
  photoacoustic reflection artifacts in ex vivo and in vivo tissue, in: 2018
  IEEE International Ultrasonics Symposium (IUS), IEEE, 2018, pp. 1--4.

\bibitem{shan2019accelerated}
H.~Shan, G.~Wang, Y.~Yang, Accelerated correction of reflection artifacts by
  deep neural networks in photo-acoustic tomography, Applied Sciences 9~(13)
  (2019) 2615.

\bibitem{reitersource}
A.~Reiter, M.~A.~L. Bell, A machine learning approach to identifying point
  source locations in photoacoustic data, in: Photons Plus Ultrasound: Imaging
  and Sensing 2017, Vol. 10064, International Society for Optics and Photonics,
  2017, p. 100643J.

\bibitem{allmansource}
D.~Allman, A.~Reiter, M.~A.~L. Bell, Photoacoustic source detection and
  reflection artifact removal enabled by deep learning, IEEE transactions on
  medical imaging 37~(6) (2018) 1464--1477.

\bibitem{caiquan}
C.~Cai, K.~Deng, C.~Ma, J.~Luo, End-to-end deep neural network for optical
  inversion in quantitative photoacoustic imaging, Optics letters 43~(12)
  (2018) 2752--2755.

\bibitem{kirchnerquant}
T.~Kirchner, J.~Gr{\"o}hl, L.~Maier-Hein, Context encoding enables machine
  learning-based quantitative photoacoustics, Journal of biomedical optics
  23~(5) (2018) 056008.

\bibitem{guttaband}
S.~Gutta, V.~S. Kadimesetty, S.~K. Kalva, M.~Pramanik, S.~Ganapathy, P.~K.
  Yalavarthy, Deep neural network-based bandwidth enhancement of photoacoustic
  data, Journal of biomedical optics 22~(11) (2017) 116001.

\bibitem{guandens}
S.~Guan, A.~Khan, S.~Sikdar, P.~Chitnis, Fully dense unet for 2d sparse
  photoacoustic tomography artifact removal, IEEE journal of biomedical and
  health informatics (2019).

\bibitem{kim2020deep}
M.~W. Kim, G.-S. Jeng, I.~Pelivanov, M.~O’Donnell, Deep-learning image
  reconstruction for real-time photoacoustic system, IEEE Transactions on
  Medical Imaging (2020).

\bibitem{gadropout}
Y.~Gal, Z.~Ghahramani, Dropout as a bayesian approximation: Representing model
  uncertainty in deep learning, in: international conference on machine
  learning, 2016, pp. 1050--1059.

\bibitem{xuphase}
Y.~Xue, S.~Cheng, Y.~Li, L.~Tian, Reliable deep-learning-based phase imaging
  with uncertainty quantification, Optica 6~(5) (2019) 618--629.

\bibitem{sahlstrom2020modeling}
T.~Sahlstr{\"o}m, A.~Pulkkinen, J.~Tick, J.~Leskinen, T.~Tarvainen, Modeling of
  errors due to uncertainties in ultrasound sensor locations in photoacoustic
  tomography, IEEE Transactions on Medical Imaging 39~(6) (2020) 2140--2150.

\bibitem{tick2019modelling}
J.~Tick, A.~Pulkkinen, T.~Tarvainen, Modelling of errors due to speed of sound
  variations in photoacoustic tomography using a bayesian framework, Biomedical
  Physics \& Engineering Express 6~(1) (2019) 015003.

\bibitem{grohl2018confidence}
J.~Gr{\"o}hl, T.~Kirchner, T.~Adler, L.~Maier-Hein, Confidence estimation for
  machine learning-based quantitative photoacoustics, Journal of Imaging 4~(12)
  (2018) 147.

\bibitem{beckfista}
A.~Beck, M.~Teboulle, A fast iterative shrinkage-thresholding algorithm for
  linear inverse problems, SIAM journal on imaging sciences 2~(1) (2009)
  183--202.

\bibitem{hojman2017photoacoustic}
E.~Hojman, T.~Chaigne, O.~Solomon, S.~Gigan, E.~Bossy, Y.~C. Eldar, O.~Katz,
  Photoacoustic imaging beyond the acoustic diffraction-limit with dynamic
  speckle illumination and sparse joint support recovery, Optics express 25~(5)
  (2017) 4875--4886.

\bibitem{vilov2020super}
S.~Vilov, B.~Arnal, E.~Hojman, Y.~C. Eldar, O.~Katz, E.~Bossy, Super-resolution
  photoacoustic and ultrasound imaging with sparse arrays, Scientific reports
  10~(1) (2020) 1--8.

\bibitem{ronnunet}
O.~Ronneberger, P.~Fischer, T.~Brox, U-net: Convolutional networks for
  biomedical image segmentation, in: International Conference on Medical image
  computing and computer-assisted intervention, Springer, 2015, pp. 234--241.

\bibitem{sridropout}
N.~Srivastava, G.~Hinton, A.~Krizhevsky, I.~Sutskever, R.~Salakhutdinov,
  Dropout: a simple way to prevent neural networks from overfitting, The
  journal of machine learning research 15~(1) (2014) 1929--1958.

\bibitem{ioffebn}
S.~Ioffe, C.~Szegedy, Batch normalization: Accelerating deep network training
  by reducing internal covariate shift, arXiv preprint arXiv:1502.03167 (2015).

\bibitem{diakogiannis2020resunet}
F.~I. Diakogiannis, F.~Waldner, P.~Caccetta, C.~Wu, Resunet-a: a deep learning
  framework for semantic segmentation of remotely sensed data, ISPRS Journal of
  Photogrammetry and Remote Sensing 162 (2020) 94--114.

\bibitem{kingmaadam}
D.~P. Kingma, J.~Ba, Adam: A method for stochastic optimization, arXiv preprint
  arXiv:1412.6980 (2014).

\bibitem{prechearly}
L.~Prechelt, Early stopping-but when?, in: Neural Networks: Tricks of the
  trade, Springer, 1998, pp. 55--69.

\bibitem{yooncc}
J.-C. Yoo, T.~H. Han, Fast normalized cross-correlation, Circuits, systems and
  signal processing 28~(6) (2009) 819.

\bibitem{wangssim}
Z.~Wang, A.~C. Bovik, H.~R. Sheikh, E.~P. Simoncelli, Image quality assessment:
  from error visibility to structural similarity, IEEE transactions on image
  processing 13~(4) (2004) 600--612.

\bibitem{schwabssim}
J.~Schwab, S.~Antholzer, R.~Nuster, M.~Haltmeier, Real-time photoacoustic
  projection imaging using deep learning, arXiv preprint arXiv:1801.06693
  (2018).

\bibitem{laksde}
B.~Lakshminarayanan, A.~Pritzel, C.~Blundell, Simple and scalable predictive
  uncertainty estimation using deep ensembles, in: Advances in neural
  information processing systems, 2017, pp. 6402--6413.

\bibitem{bench2020towards}
C.~Bench, A.~Hauptmann, B.~Cox, Towards accurate quantitative photoacoustic
  imaging: learning vascular blood oxygen saturation in 3d, arXiv preprint
  arXiv:2005.01089 (2020).

\end{thebibliography}
                            
\end{document}


\begin{frontmatter}

\title{Compensating for visibility artefacts in photoacoustic imaging with a deep learning approach providing prediction uncertainties: supplementary materials}

\author[lip]{Guillaume Godefroy\corref{email1}}
\author[lip]{Bastien Arnal}
\author[lip]{Emmanuel Bossy}
\address[lip]{Univ. Grenoble Alpes, CNRS, LIPhy, 140 rue de la Physique, CS 40700, 38058 Grenoble CEDEX 9, FRANCE}                                             
\cortext[email1]{Corresponding author: guillaume.godefroy@univ-grenoble-alpes.fr}

\begin{abstract}                          
This document provides supplementary information to "Compensating for visibility artefacts in photoacoustic imaging with a deep learning approach providing prediction uncertainties". Included are a schematic representation of the deep learning algorithm (DLA), a comparison of the prediction of the DLA based on either demodulated beamformed (dmBF) image or modulated beamformed (mBF) image as input, the prediction from experimental data for a DLA trained with simulation data and the uncertainty estimation of the instances presented in the main text. The reconstruction of a simulated vessel-like object, provided by the DLA trained on the simulated leaf dataset, is also presented.
\end{abstract}

\begin{keyword}                           
Photoacoustic imaging; Deep learning; Visibility artefacts; Monte Carlo dropout; Bayesian neural network    \end{keyword}  

\end{frontmatter}

\section{Network architecture}
\begin{figure}[H]
\centering
\includegraphics[scale=0.5]{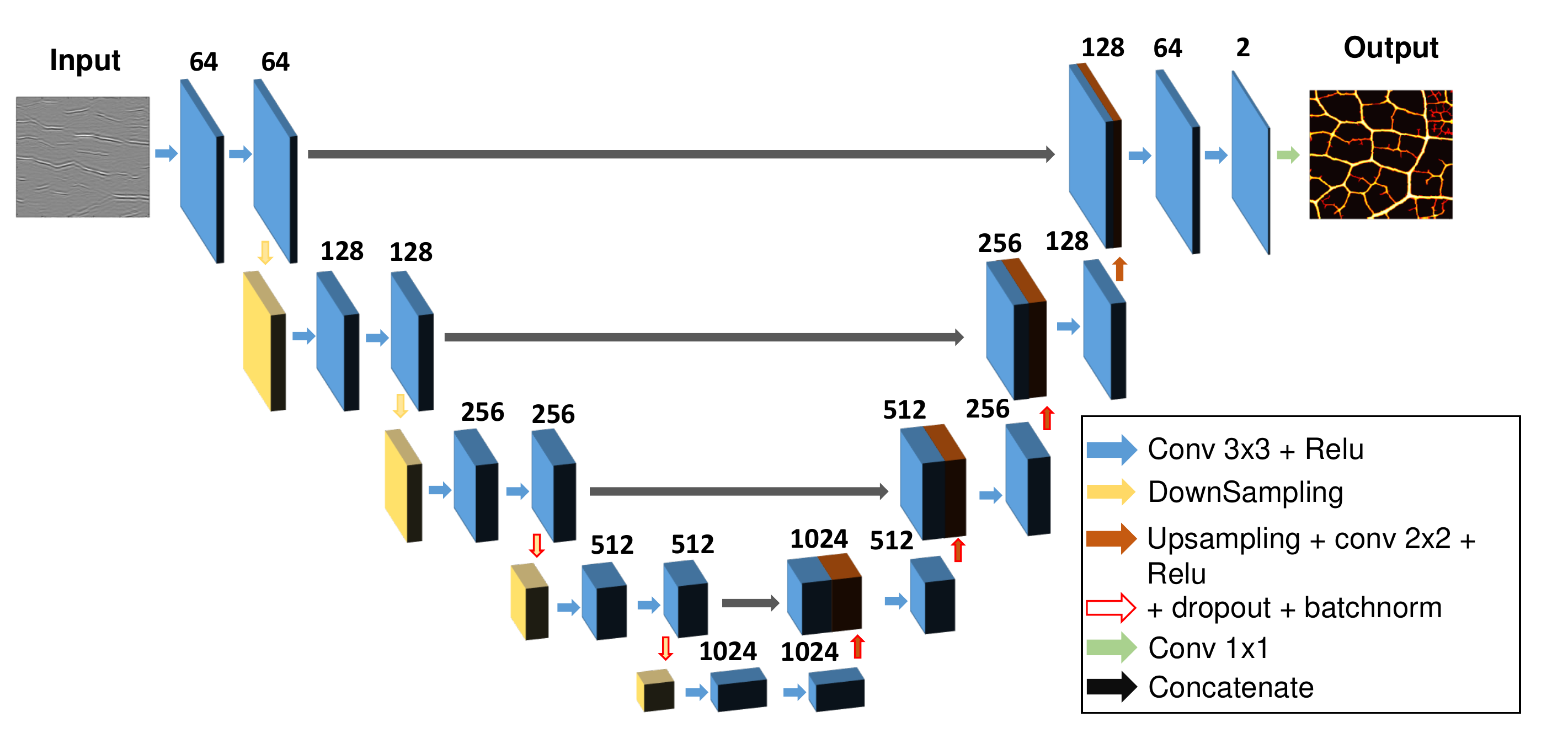}
\caption{Network architecture.}
\label{figs1}
\end{figure}

Unet is a well known network architecture first developed for segmentation task. Our implementation is shown in Fig. \ref{figs1}. It is a convolutional neural network composed of two paths: the contracting and expanding path. The first one, called the encoder, is a traditional stack of convolutional and pooling layers where the network extracts more and more complex features. The second one, called the decoder, is the symmetric expanding path where pooling operations are replaced by  upsampling operators to recover at the output the resolution of the input. Context information is propagated from the encoder to the decoder through skip connections to provide local information to the global information while upsampling (black arrows). The weights are initialized with samples from a truncated normal distribution centered on 0 with standard deviation depending of the number of units in the weight tensor.
Dropout layers are added to this architecture. Dropout is a popular regularization technique to limit overfitting. A certain set of neurons, chosen randomly, are disabled at each training step. This prevents units from co-adapting too much and forces the network to learn more robust features. Batch normalization  normalizes the output of the previous activation layer by subtracting the batch mean and dividing by the batch standard deviation. It helps to speed up the learning and also reduces overfitting by adding some noise, similarly as dropout.

\section{mBF or dmBF image as input of the network}
\begin{figure}[H]
\centering
\includegraphics[scale=0.6]{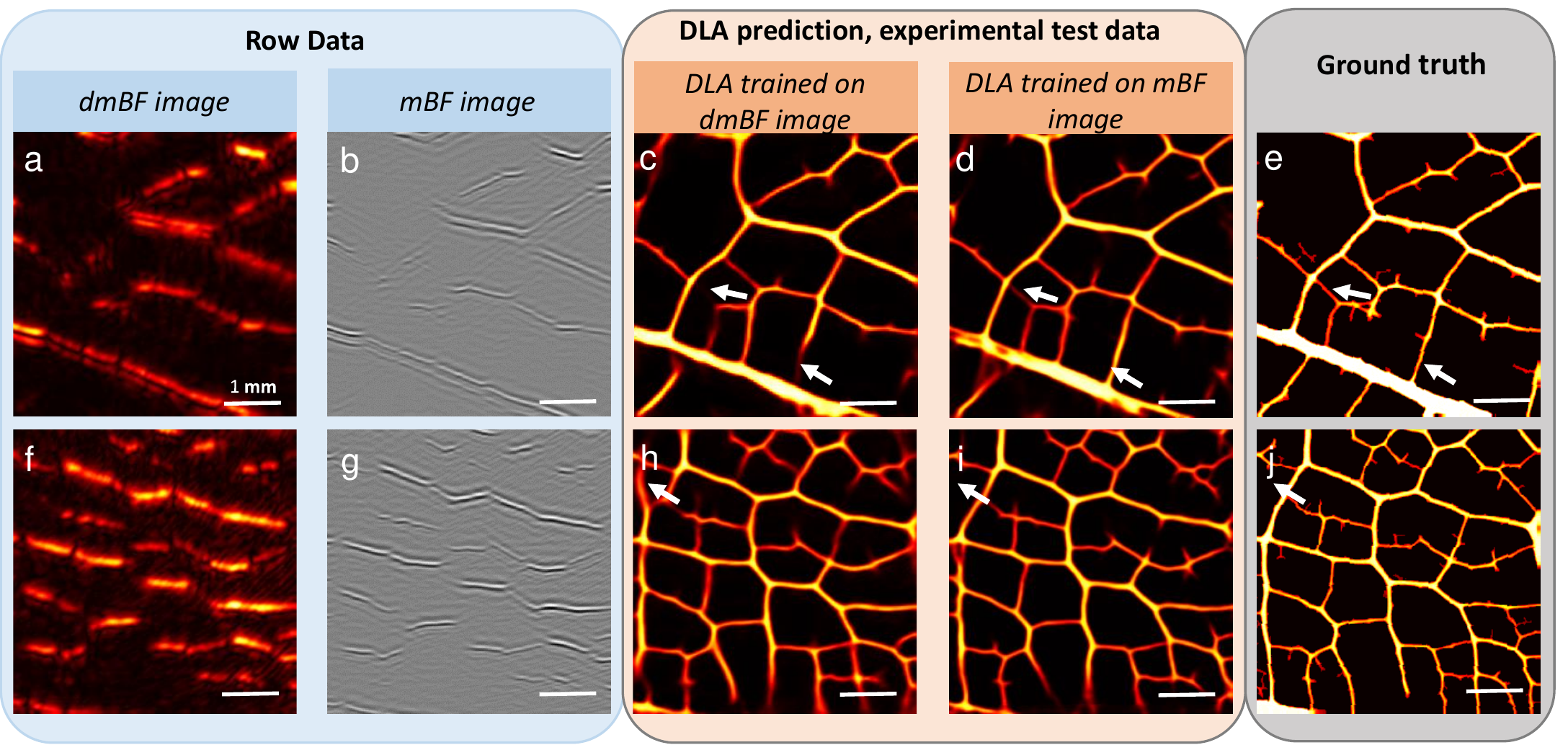}
\caption{Deep learning algorithm (DLA) prediction on experimental data for dmBF PA image or mBF PA image as input, two examples.
\newline
\textbf{a, f, }dmBF image.\textbf{ b, g, }mBF image.\textbf{ c, h, }Prediction with DLA trained on dmBF data.\textbf{ d, i, }Prediction on experimental data with DLA trained on mBF image.\textbf{ e, j, }Ground truth.
}
\label{figs2}
\end{figure}

The input of the network is obtained from the delay and sum algorithm (DAS) applied to time signals. When applied to real time signals, DAS provides a mBF image. When applied to complex signals obtained with a Hilbert transform, DAS provides a complex image whose modulus is the dmBF image. The mBF image and the dmBF image are the two types of input that we consider here. 
One DLA was trained for each type of input, the corresponding predictions are displayed in Fig. \ref{figs2}. The algorithm performs better on the mBF image, leading to a scaled and shifted structured similarity index (sSSIM) of 0.76 instead of 0.72. The prediction from the dmBF image suffers from more artefacts (arrows) and the DLA fails to recover the true vessels thicknesses, which are over estimated. The mBF image, despite being more different from the true physical structure of the object than  the dmBF image, and thus from the ground truth, carries more information to be captured by the network.

\section{Reconstruction of experimental data with DLA trained on simulation data}

\begin{figure}[H]
\centering
\includegraphics[scale=0.7]{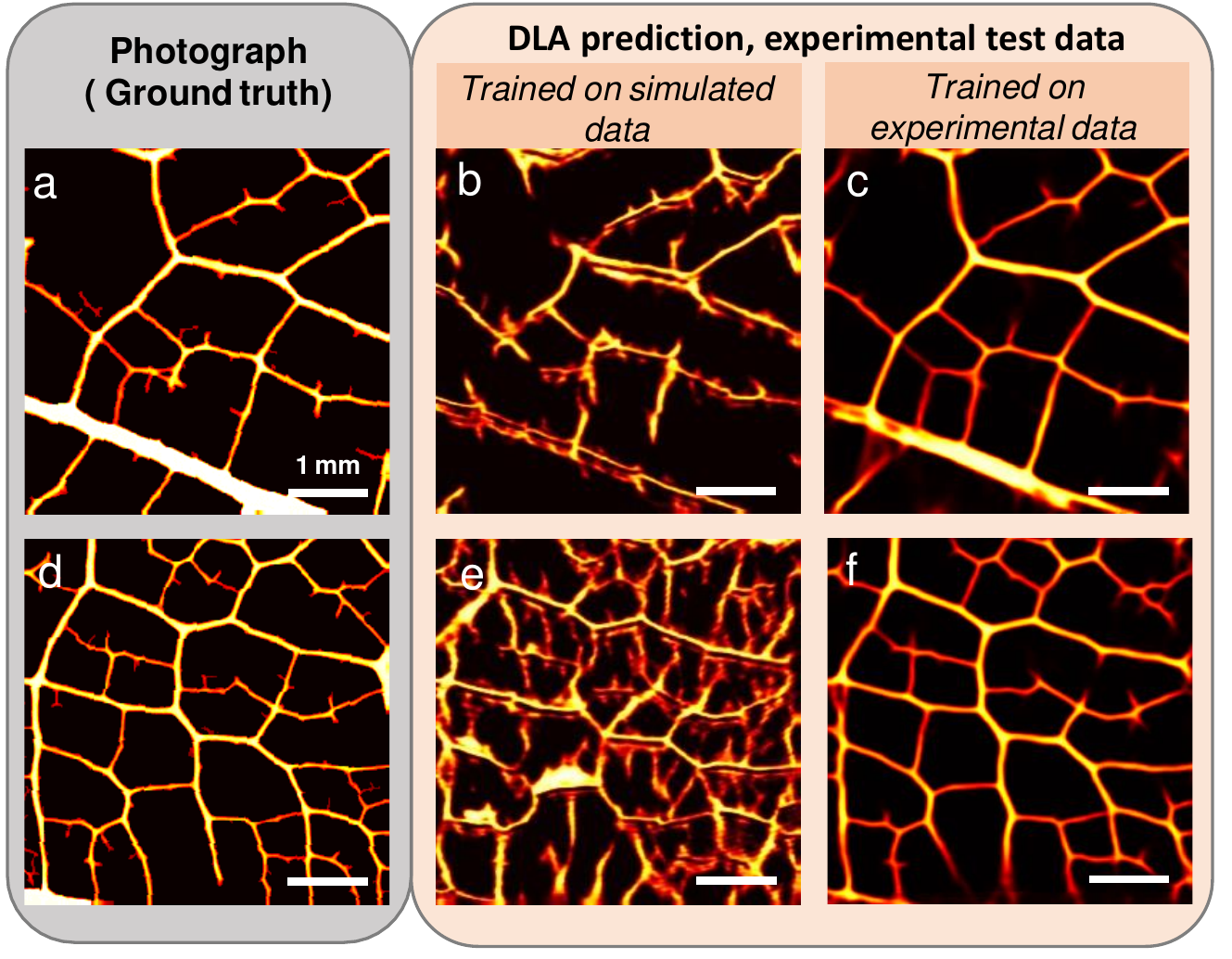}
\caption{Deep learning algorithm (DLA) prediction on experimental data for DLA trained on simulated data and DLA trained on experimental data, two examples. \textbf{a, d, }Ground truth.\textbf{ b, e, }Prediction with DLA trained on simulated data.\textbf{ c, f, }Prediction on experimental data with model trained on experimental data.}
\label{figs3}
\end{figure}

Predictions from neural networks trained with simulation datasets or experimental datasets are presented in Fig. \ref{figs3}. Although the DLA trained on simulation data still manages to find several vertical structures that are not visible on the DAS image, the predicted image is polluted by a lot of artefacts. Clearly, experimental data are necessary to train efficiently the model. However, as shown in the main text, pretraining the network on a simulation dataset allows reducing the size of the experimental training set.  

\section{Uncertainty estimation}
\begin{figure*}[ht]
\centering
\includegraphics[scale=1.00]{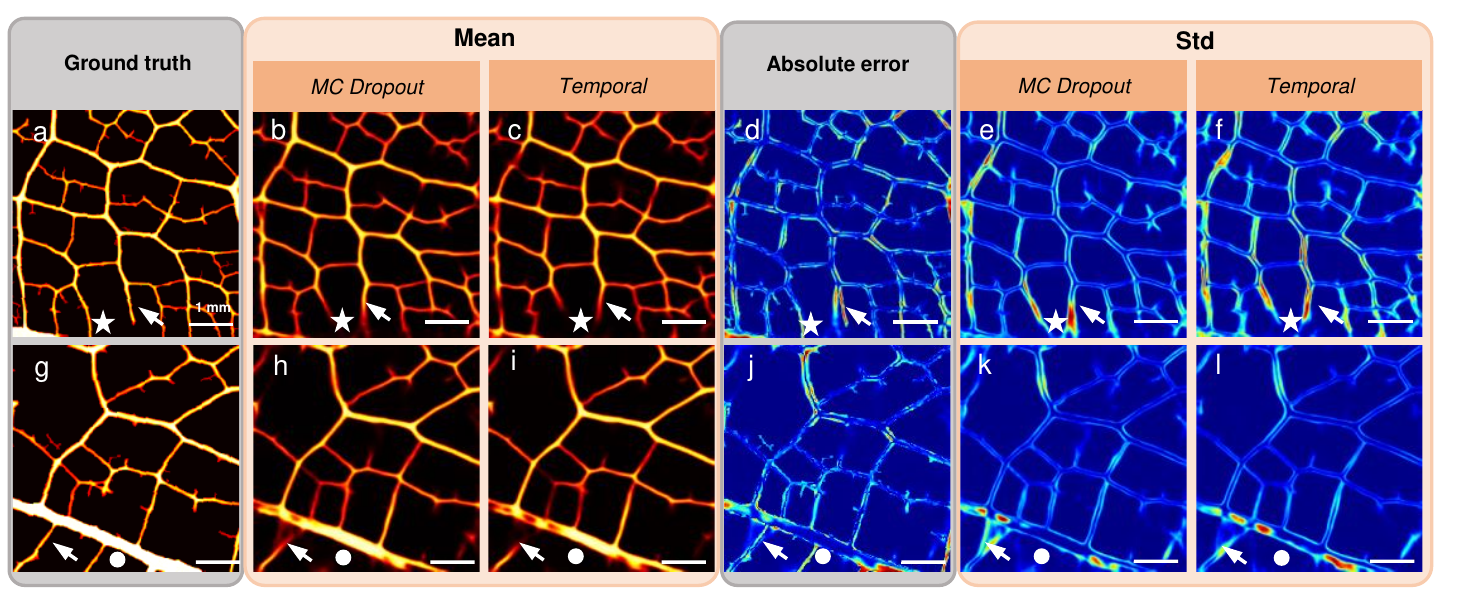}
\caption{Uncertainty estimation on experimental mBF image.
\textbf{ a, g, }Ground truth: photograph of the object.\textbf{ b, h, }Mean image of the object computed over 20 inferences generated from a unique acquisition with the deep learning algorithm, dropout activated.\textbf{ e, k, }Corresponding STD.\textbf{ c, i, }Mean image of the object computed over prediction generated from several RF signals acquired at different time with the deep learning algorithm, dropout disabled.\textbf{ f, l, }Corresponding STD\textbf{ d, j, }Absolute error between the ground truth (a,g) and the mean (b,h).
}
\label{figs4}
\end{figure*}

Uncertainty estimation of the two previous examples are presented in Fig. \ref{figs4}. Similarly to the example in the main text, the standard deviation (std) map helps to locate errors in the reconstruction such as invented structures and incorrect orientation or position  (arrows). One can notice that although the value of the std is higher at location of some missing veins (star), some of them are not even displayed on the std (circle).
The lack of information in the data may be more important for these structures, misleading the DLA.

\section{Generalization ability}

\begin{figure*}[ht]
\centering
\includegraphics[scale=.8]{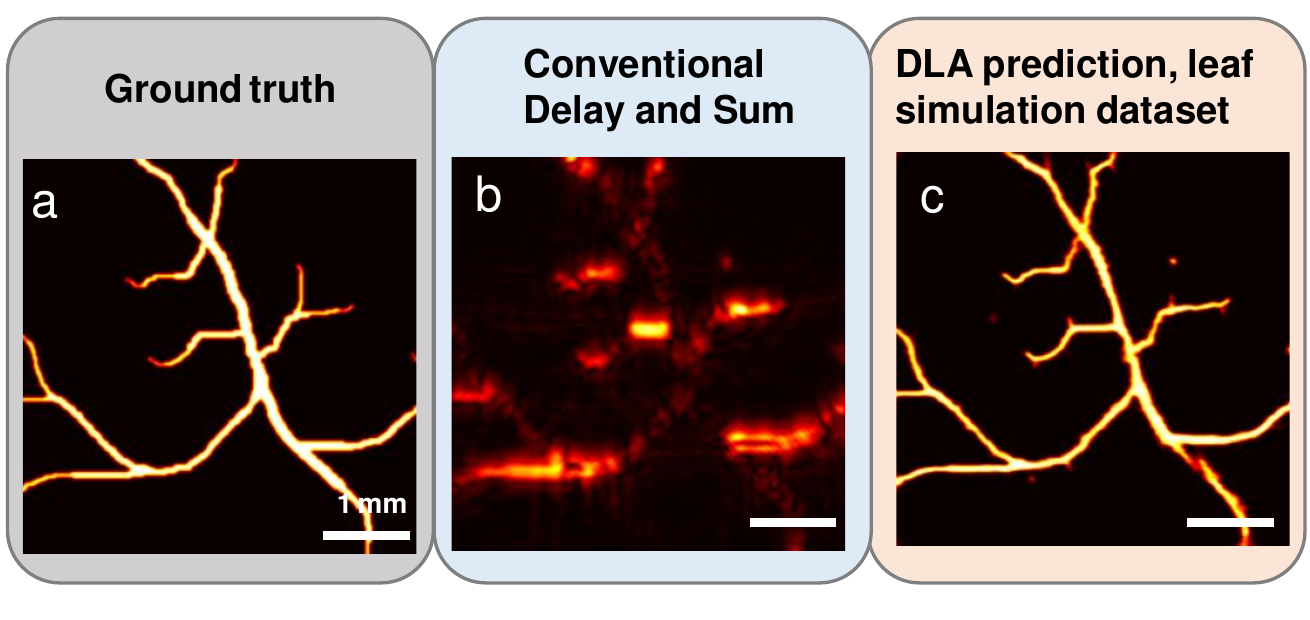}
\caption{Reconstruction of a simulated vessel-like objet with a network trained on simulation from the leaf object.
\textbf{a, }Ground truth. \textbf{b, }dmBF image, delay and sum. \textbf{c, }Prediction of the deep learning algorithm.
}
\label{figs5}
\end{figure*}

This work was mostly dedicated to demonstrate the capacity of a neural network to correct visibility artefacts on a specific class of object. However the possibility or reconstructing objects from another class is an open question. To illustrate this problematic on a specific example, we used the network trained on leafs (simulated signals) to reconstruct the PA image of a vessel-like structure (test image from the k-wave package, http://www.k-wave.org/). The result is presented in Fig. \ref{figs5}. Although the class of the object is clearly different, the algorithm turns out to perform extremely well on this specific example, as visibility artefacts are removed, suggesting no or low overfitting. However, this is just one example, obtained from simulated data, and further studies are required to investigate the generalization limits of the network, especially for experimental data.

